\documentclass[conference]{IEEEtran}
\usepackage{amsmath,graphicx,multirow,slashbox,algpseudocode,subfigure,threeparttable}

\ifCLASSINFOpdf
\else
\fi

\begin{document}
%
\title{PipeCNN: An OpenCL-Based FPGA Accelerator for Large-Scale Convolution Neuron Networks}

\author{\IEEEauthorblockN{Dong Wang, Jianjing An and Ke Xu}
\IEEEauthorblockA{Institute of Information Science\\
Beijing Jiaotong University\\
Beijing 100044, China\\
Email: wangdong@bjtu.edu.cn}
}

\maketitle

\begin{abstract}
Convolutional neural networks (CNNs) have been widely employed in many applications such as image classification, video analysis and speech recognition. Being compute-intensive, CNN computations are mainly accelerated by GPUs with high power dissipations. Recently, studies were carried out exploiting FPGA as CNN accelerator because of its reconfigurability and energy efficiency advantage over GPU, especially when OpenCL-based high-level synthesis tools are now available providing fast verification and implementation flows. Previous OpenCL-based design only focused on creating a generic framework to identify performance-related hardware parameters, without utilizing FPGA's special capability of pipelining kernel functions to minimize memory bandwidth requirement. In this work, we propose an FPGA accelerator with a new architecture of deeply pipelined OpenCL kernels. Data reuse and task mapping techniques are also presented to improve design efficiency. The proposed schemes are verified by implementing two representative large-scale CNNs, AlexNet and VGG on Altera Stratix-V A7 FPGA. We have achieved a similar peak performance of 33.9 GOPS with a 34$\%$ resource reduction on DSP blocks compared to previous work. Our design is openly accessible and thus can be reused to explore new architectures for neural network accelerators.
\end{abstract}

\IEEEpeerreviewmaketitle

\section{Introduction}
\label{sec:intro}
Convolutional neural network (CNN), as an emerging deep learning architecture, has received huge attentions in various applications, such as video surveillance, image searching, speech recognition, and robot vision. A CNN works with multiple convolution layers that extract features from input data, followed by classification layers making decisions. Typical large-scale CNNs \cite{AlexNet,Vgg} usually consist of millions of neural units and millions of connections that require over billion operations to process only one input.

General purpose CPUs, being sequential systems with limited computational resources, are inefficient for implementing CNN-based compute-intensive applications. Currently, GPUs are widely adopted as hardware accelerators for training deep neuron networks. However, they are energy inefficient for embedded applications. FPGAs, which provide massive processing elements, reconfigurable interconnections and lower power dissipation, are naturally suitable to implement neural network circuits.
Studies, such as \cite{DLAU2016}, \cite{Zynq2016} have reported efficient CNN accelerators on embedded FPGA platforms. However, traditional register-transfer-level (RTL) design flow adopted in these studies require
great effort in writing complex RTL codes, practicing time-consuming simulations and compilations before one can actually run accelerators on hardware.

High-Level Synthesis (HSL) tools, which enable automatic compilation from high-level program (C/C++) to low-level RTL specifications, were recently adopted by many studies to implement deep neural networks on FPGAs. In \cite{FPGA2015}, an accelerator design was implemented by using the Vivado-HLS tool on a Xilinx VC707 FPGA. Computation throughput and memory bandwidth are quantitatively explored by using a roofline model to find the design with the best performance and lowest resource. However, only convolution layers are implemented. The work of \cite{FPGA2016} proposed a fixed-point CNN design using the OpenCL framework. A systematic methodology is presented to minimize execution time with given resource constraints. Due to a matrix multiplication-based kernel design for convolution layer and the GPU-like separated kernel organization adopted, FPGA's special advantage of implementing deeply pipelined circuits (kernels) is not fully exploited to better improve computation throughput and minimize memory bandwidth.

Main contribution of this work are: (1) an OpenCL-based FPGA accelerator with an efficient structure of pipelined kernels is proposed for implementing large-scale CNNs; (2) the design space of the proposed architecture was fully explored on Stratix-V A7 FPGA and two real-word large-scale CNN models were implemented and tested. Results show that the proposed scheme achieves improved performance and resource utilization than previous works; (3) we have made our design openly accessible \cite{github} for other researchers to study and explore new accelerator architectures for deep neural networks.

%

\section{OpenCL-Based CNN Implementation}

\subsection{OpenCL Framework}

\begin{figure}[t]
\begin{center}
 \includegraphics[width=2.4in]{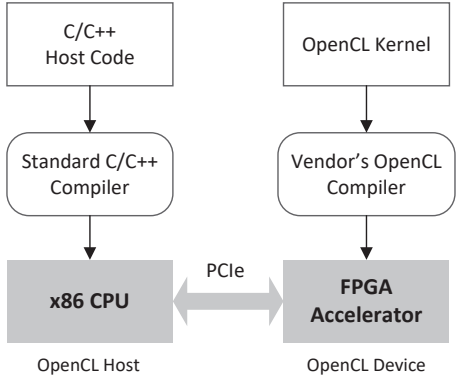}
\caption{OpenCL-based FPGA design flow for CNN accelerator.}
\label{fig:flow}
\end{center}
\end{figure}

OpenCL is an open, cross-platform parallel programming language that can be used in both GPU and FPGA developments. The OpenCL-based FPGA accelerator development flow is summarized in Fig.~\ref{fig:flow}. In the framework, an FPGA board (as OpenCL device) is connected with a desktop CPU (as OpenCL host) through a high speed PCIe slot forming a heterogenous computing system. An OpenCL code, which defines multiple parallel compute units (CUs) in the form of kernel functions, is compiled and synthesized to run on the FPGA accelerator. On the host side, a C/C++ code runs on the CPU, providing vendor specific application programming interface (API) to communicate with the kernels implemented on the FPGA accelerator. This work uses the Altera OpenCL SDK toolset for compiling, implementing and profiling the OpenCL codes on FPGAs.

\subsection{Proposed Accelerator Architecture}
A standard CNN  \cite{AlexNet, Vgg} for image classification is comprised of one or more convolutional layers, pooling layers, followed by one or more fully connected (FC) layers. As analyzed in \cite{FPGA2016}, the core part of the convolution layer is a 3-dimensional multiply-accumulate operation that can be defined by

\begin{scriptsize}
\begin{equation}
D_o(f_o,y,x)= \sum^{C_l}_{f_i=0}\sum^{K}_{k_y=0}\sum^{K}_{k_x=0}W_l(f_o,f_i,k_y,k_x) \cdot D_i(f_i,y+k_y,x+k_x)
\label{func:conv}
\end{equation}
\end{scriptsize}
where $D_i(f_i,y,x)$ and $D_o(f_o,y,x)$ denote the neurons at position $(x,y)$ in the input feature map $f_i$ and output feature map $f_o$, respectively.
$W_l(f_o,f_i,y,x)$ represents the corresponding weights in the $l$-th layer that gets convolved with $f_i$.
In pooling layers, 2-D subsampling operations are performed on neighboring neurons of the same feature map $f_i$. As traversing deeper in the neural network, feature dimensions are gradually reduced. In FC layers, each output neuron is calculated by the weighted summation of all input neurons shown by

\begin{scriptsize}
\begin{equation}
D_o(f_o)= \sum^{C_l}_{f_i=0}W_l(f_o,f_i) \cdot D_i(f_i)
\label{func:fc}
\end{equation}
\end{scriptsize}
In some CNN models \cite{AlexNet}, local response normalization (LRN) layers that perform normalization operations on each input neuron value by a factor that depends on the neighboring neurons are also used following the pooling layer.

As illustrated in Fig.~\ref{fig:top-arch}, the proposed architecture consists of four kernels that are connected by using Altera's OpenCL extension Channel/Pipes. The Convolution kernel (Conv.) is designed to implement both the 3-D multiply-accumulate operation of (\ref{func:conv}) and the inner product operation of (\ref{func:fc}). The Pooling kernel performs the subsampling directly on the output data streams of the Conv. kernel. Two data mover kernels, namely MemRD and MemWR, transfer feature data and weights from/to the global memory. As analyzed in Section~\ref{sec:intro}, the cascaded kernels form a deep computation pipeline that can implement a serial of basic CNNs operations without the need of storing interlayer data back to global memory. It significantly reduces the bandwidth requirement compared to the work of \cite{FPGA2016}. The LRN function is implemented separately from the pipeline since it may function on data from adjacent feature maps or the same feature map, which requires multiple memory access patterns. Detailed design of each kernel is as follow:

\begin{figure}[t]
\begin{center}
 \includegraphics[width=3.3in]{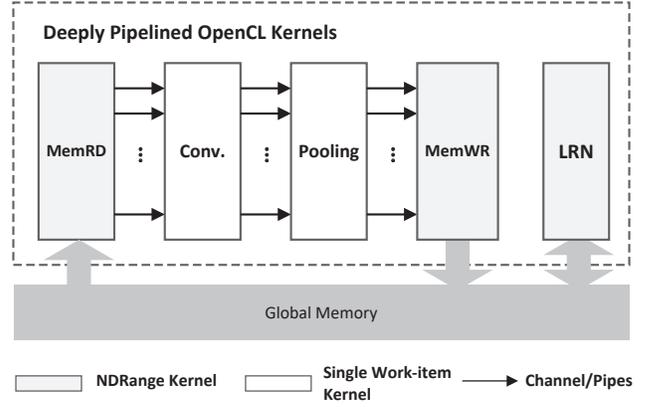}
\caption{The Proposed Architecture of the CNN Accelerator.}
\label{fig:top-arch}
\end{center}
\end{figure}

\begin{figure}
\begin{algorithmic}[1]
\State Define shift-register $Reg[N]$ as delayed buffer
\State Set required convolution counter bound $CN$
\State \emph{\#pragma unroll by factor of $CU\_NUM$}
\For {each convolution pipeline}
\For {each output neuron $D_o$}
\State Initializing all $Reg[n]$ to zeros
\For {$j=0; j<CN$}
\State Read vectorized $D_i(j)$ from Channel
\State Read vectorized $W_l(j)$ from Channel
\State Perform parallel multiply-add operation of
\State $Temp=W_l(j)\times D_i(j) + Reg[N]$
\State Perform register shifting $Reg[n]=Reg[n-1]$
\State Store result $Temp$ in $Reg[0]$
\EndFor
\State Perform parallel summation $D_o= \sum Reg[n]$
\EndFor
\EndFor
\end{algorithmic}
\caption{Pseudo-code of the convolution kernel.}
\label{fig:code-conv}
\end{figure}

\subsubsection{Convolution Kernel}
A single-threaded kernel with parallel convolution data paths is designed to implement both the functions of the convolution and FC layers. Two techniques are used to improve the computation throughput and pipeline utilization. First, a multi-mode convolution circuit with a structure of deeply pipelined multiply-add tree and delayed buffers is designed. In \cite{FPGA2015}, Eq.~(\ref{func:conv}) was written as a 5-level nested loop, in which complicated loop tiling and memory partition techniques are used to improve computation throughput. However, manual memory partition capability is not yet available in Altera's OpenCL and the Channel read/write operations used in loops will also prevent tiling optimization. Therefore, we transform (\ref{func:conv}) into a structure similar to (\ref{func:fc}), and implement both functions as a 2-level nested loop structure. The pseudo-code is shown in Fig.~\ref{fig:code-conv}. When the appropriate buffer depth \emph{N} is set, an efficient pipeline with an initial interval of two can be synthesized by Altera's OpenCL compiler.

Secondly, data vectorization and parallel CU structures are both exploited in the design. Vectorized input features $D_i$ and weights $W_l$ are streamed by multiple Channels. A design parameter \emph{VEC\_SIZE} is introduced to control the input throughput. The outermost \emph{for} loop is unrolled by a factor of \emph{CU\_NUM} to generate multiple instances of the convolution pipeline. Consequently, outputs $D_o$ in different output feature maps $f_o$ can be generated in parallel. When configured in 3D convolution mode, $CN$ is set to the value of $K\times K\times C'$, while in FC mode, $CN$ is set to $C'$, where \emph{C'=C/VEC\_SIZE}. When no pipeline stall are caused by Channel access, a speedup by \emph{VEC\_SIZE} $\times$ \emph{CU\_NUM} can be achieved.


\subsubsection{Data Mover Kernels}
\begin{figure}[t]
\begin{center}
 \includegraphics[width=3.5in]{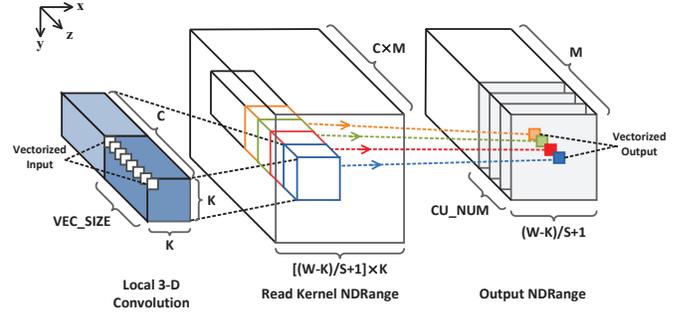}
\caption{Data and work-item mapping scheme of the data mover kernels.}
\label{fig:map}
\end{center}
\end{figure}

Two multi-mode 3D NDRange kernels are designed to fetch/store data from/to the global memory for the computation pipelines. Data and work-item mapping schemes are illustrated in Fig.~\ref{fig:map}. In convolution mode, the MemRd kernel launches with a global work size of $([(W-K)/S+1]\times K, [(W-K)/S+1]\times K, C' \times M)$, while the MemWR kernel works in a NDRange of $((W-K)/S+1, (W-K)/S+1, M)$. Work-items are arranged into multiple concurrent work-groups, each of which has a local work size of \emph{(K, K, C')}. Therefore, a strict data read-write ratio is assigned to the two kernels.
The proposed data movers enable efficient data reuses which can significantly reduce the global memory bandwidth requirements: 1) for each work-item, the fetched data $D_i(f_i,y,x)$ of the input feature map $f_i$ are replicated by registers inside the MemRD kernel, and then passed to all the CUs of the following Conv. kernel for parallel computation of the \emph{CU\_NUM} output features.
2) the fetched weights $W_l$ are first loaded onto an on-chip cache generated by the compiler. Different work-groups that share the same work-group index $z$ can reuse the same weights from the cache without issuing new global memory load instructions.

In FC mode, both the input feature and weight data are 1D vectors as defined in Eq.~(\ref{func:fc}). Directly launching MemRD kernel with only one classification task will significantly reduce the opportunity of data reuse in weights. Therefore, we introduce batched processing capability in MemRD. For instance, a batch of $64$ classifications can be processed with a single kernel launch by mapping the input feature maps as a single 3D data set with the setting of $(C, 8, 8)$.


\subsubsection{Pooling Kernel}
\begin{figure}[t]
\begin{center}
 \includegraphics[width=2.8in]{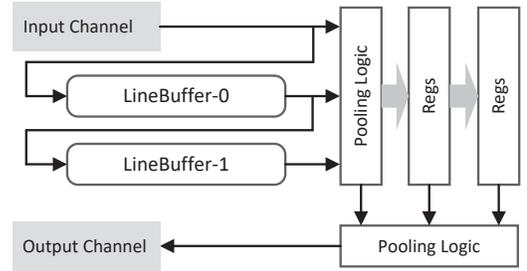}
\caption{Line buffer-based hardware architecture of the pooling kernel ($L=2$).}
\label{fig:pool}
\end{center}
\end{figure}

A line-buffer-based hardware structure is proposed for the pooling kernel as shown in Fig.~\ref{fig:pool}. The kernel first reads data of the same feature maps in a line-by-line manner from the Channel and then store them in a group of $L$ line buffers. After all buffers are fully filled up, a window of feature map data are read out and send to the next stage of pooling logic. In CNNs, two pooling schemes, i.e., max-pooling and average-pooling, are widely used. Therefore, the pooling logic modules support selecting the maximum or computing the average value of the \emph{(L+1)} inputs. The kernel can also be turned off by setting a control register.

\subsubsection{LRN Kernel}
We choose the piece-wise linear approximation scheme presented by \cite{FPGA2016} to implement the core exponent function of the LRN kernel. We improve this scheme by introducing a new lookup table segmentation scheme to reduce the hardware costs as shown in Fig.~\ref{fig:code-lrn}. In this new method, we divide the function evaluation range by using power of $2^{-n}$, where $n$ is a integer that controls the accuracy. The approach avoids complicated table addressing logic by directly operates on the exponent of the input. The hardware parameter $Shif\_Bit$ is determined by the used segmentation parameter $n$. In AlexNet implementation, a maximum approximation error of $0.5\%$ is achieved by setting $n=2$.

\begin{figure}
\begin{algorithmic}[1]
\State Load the feature values into a local memory $FIN[i,j]$
\State Place barrier on local memory $FIN[i,j]$
\For {each $i,j$}
\State Perform parallel access on $FIN[i,j]$ to fetch all
\State neighboring features $Features[p,q]$
\State Compute $Sum\_of\_Features = \sum{Features[p,q]}$
\State Copy the exponent $Exp$ from $Sum\_of\_Features$
\State Set $Addr = Exp \gg Shift\_Bit + 1$
\State Access look-up table by $Addr$
\State Compute the approximated function value $pwlf$
\State Compute result $FOUT[i,j]=FIN[i,j]*pwlf$
\EndFor
\State Place barrier on local memory $FIN[i,j]$
\State Store $FOUT[i,j]$ back to global memory
\end{algorithmic}
\caption{Pseudo-code of the LRN kernel. pwlf refers to a piece-wise linear approximation operation.}
\label{fig:code-lrn}
\end{figure}

\begin{figure*}[t]
\begin{minipage}{1.0\linewidth}\centering
\subfigure[]{\includegraphics[width=1.7in]{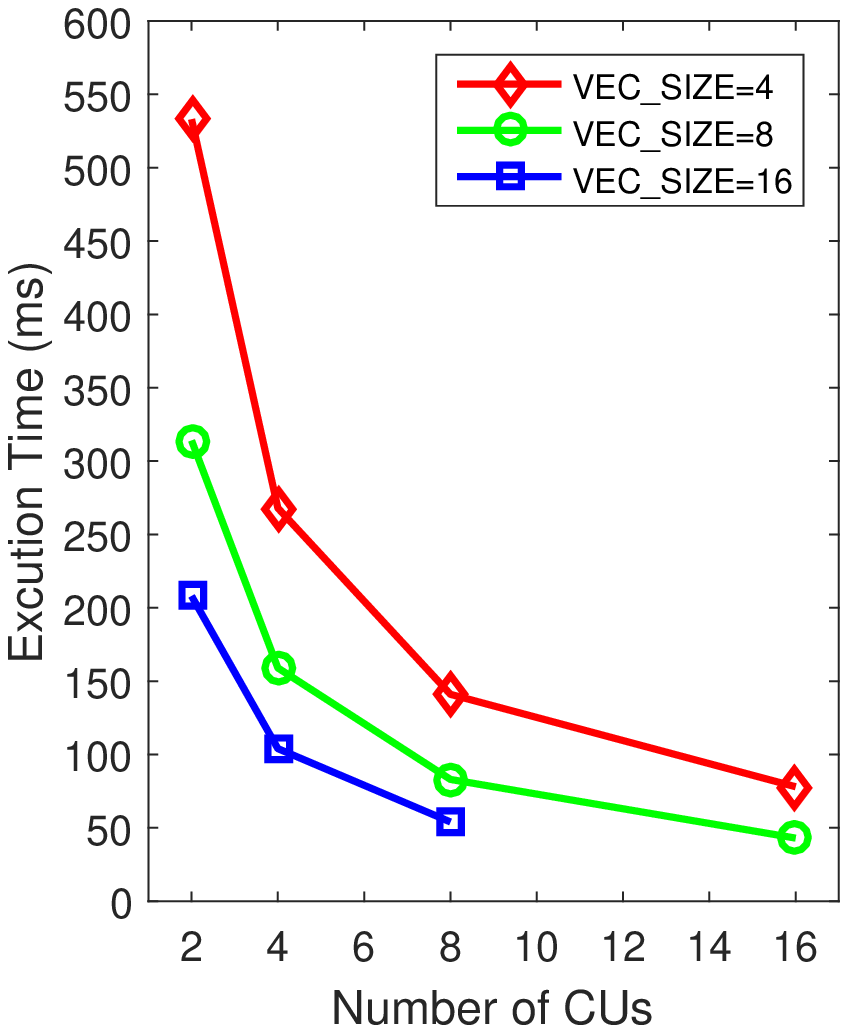}}
 \subfigure[]{\includegraphics[width=1.8in]{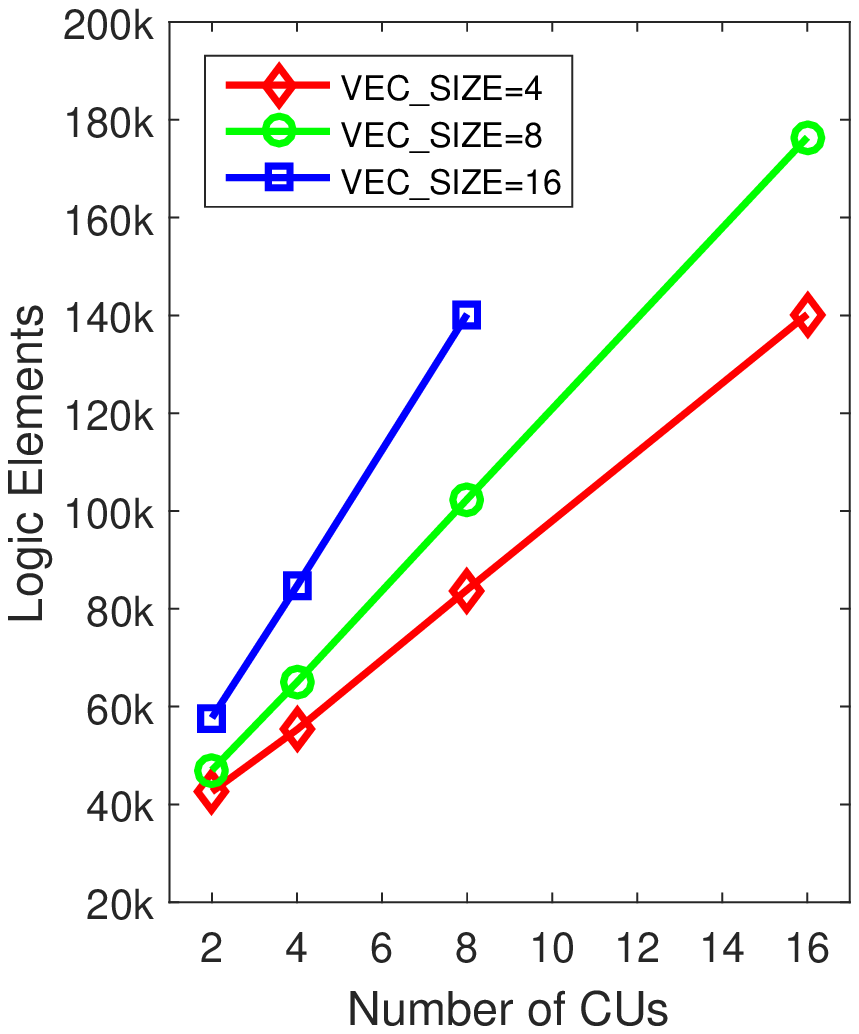}}
 \subfigure[]{\includegraphics[width=1.7in]{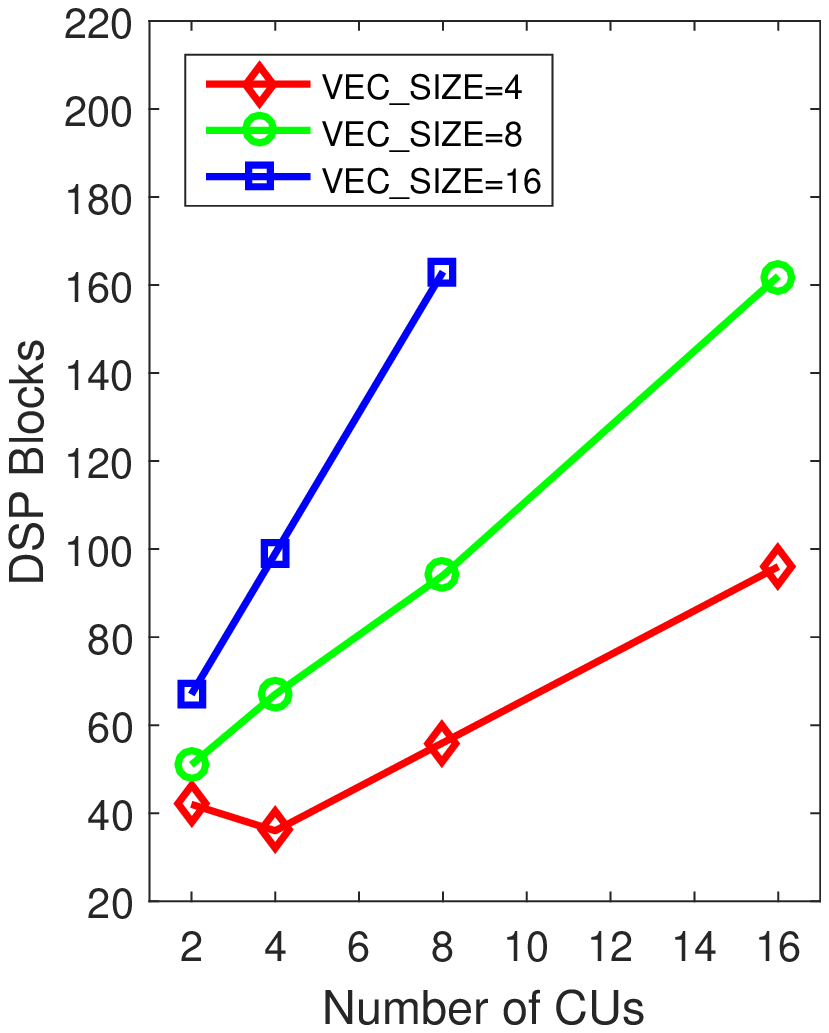}}
  \subfigure[]{\includegraphics[width=1.7in]{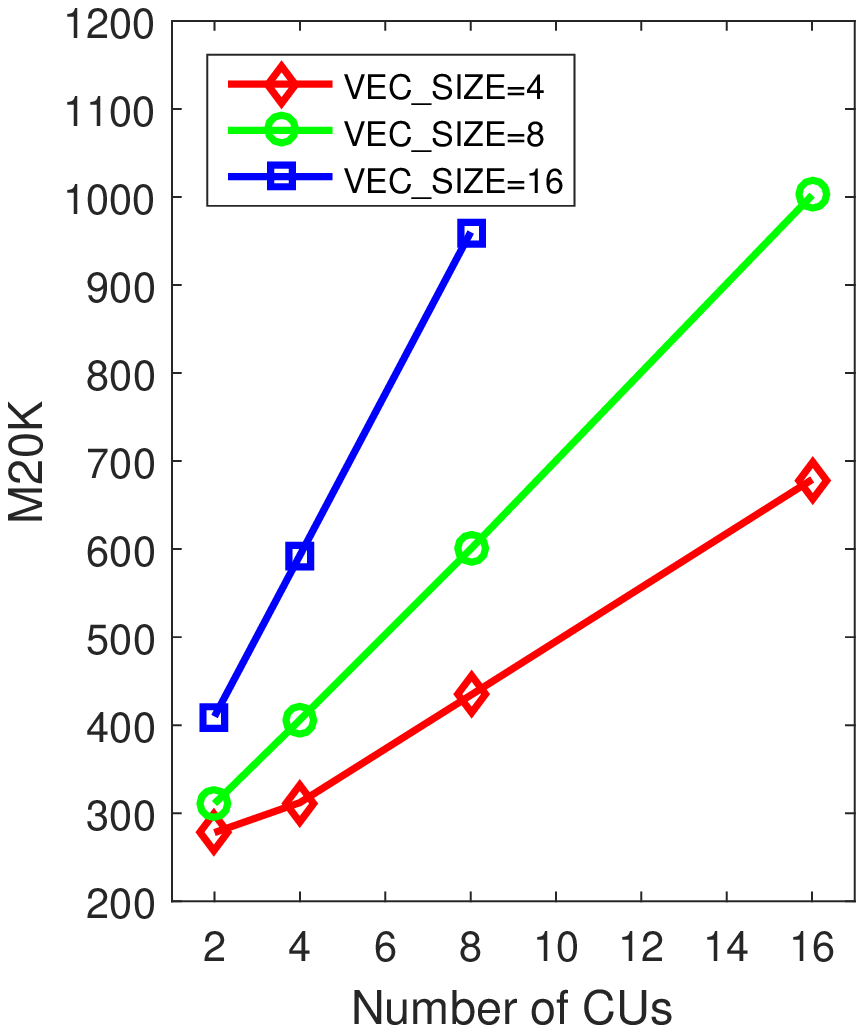}}
\caption{Design space exploration for AlexNet model on DE5-net FPGA board. The design with parameters \emph{VEC\_SIZE=16} and \emph{CU\_NUM=16} is too large to fit in the FPGA device, and is not reported.}
 \label{fig:analyze}
\end{minipage}
\end{figure*}

\begin{figure}[t]
\begin{minipage}{1.0\linewidth}\centering
\subfigure[]{\includegraphics[width=3.5in]{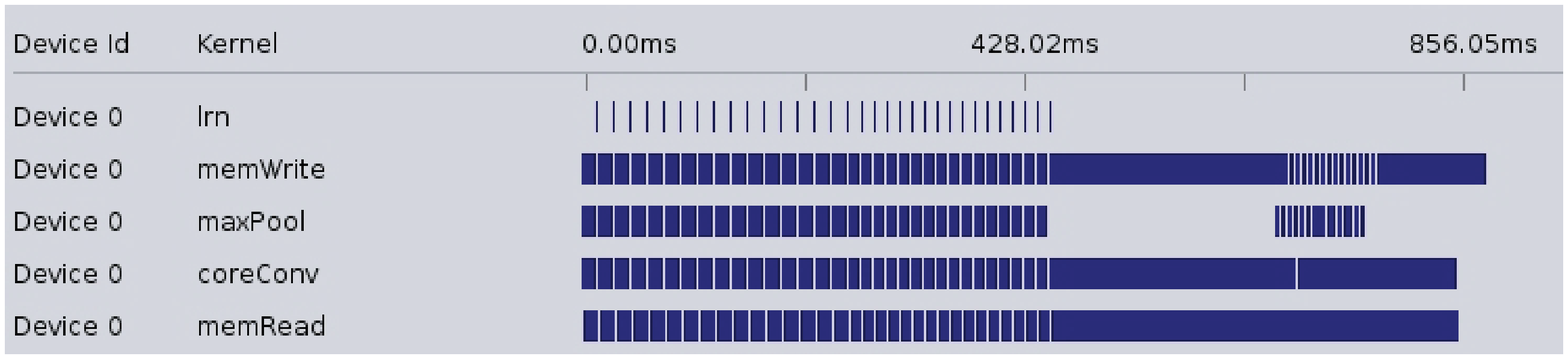}}
 \subfigure[]{\includegraphics[width=3.5in]{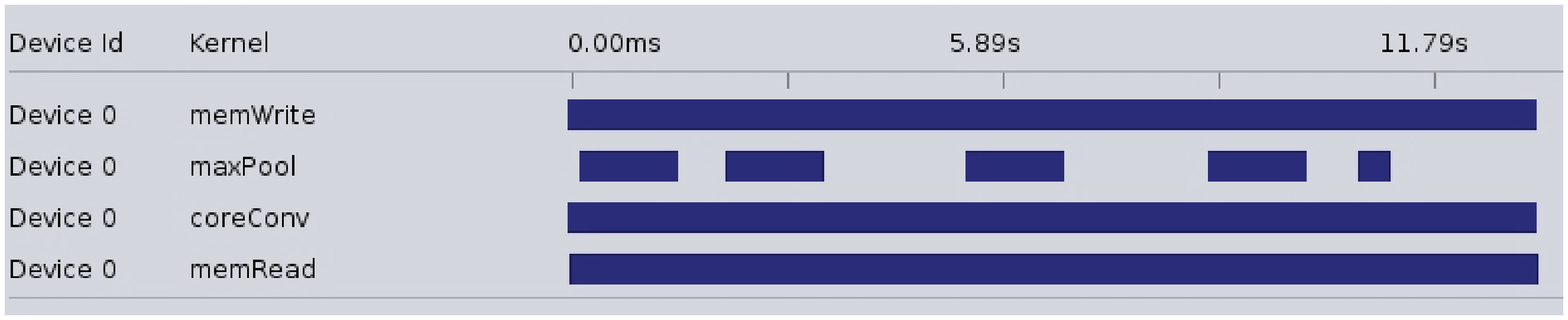}}
\caption{The profiled execution timeline of the OpenCL kernels running (a) AlexNet and (b) VGG-16 models. The configured batch size is 16.}
 \label{fig:timeline}
\end{minipage}
\end{figure}

\section{Experimental Results}
In this section, we present the implementation results of the proposed OpenCL design on Altera Stratix-V FPGA based DE5-net board. The Stratix-V A7 FPGA consists of 622K logic elements (LEs), 256 DSP blocks and 2560 M20K RAMs. There are also two 2GB DDR3 DRAMS connected to the FPGA that function as the global memory.
The OpenCL kernel codes are compiled by using Altera OpenCL SDK v15.1. Two large-scale CNN models: AlexNet (8 layers) and VGG (16 layers) models are tested with different hardware parameter settings to explore the design space.

Fig.~\ref{fig:timeline} presents the measured performance (a) and detailed resource utilization (b-d) of the FPGA for implementing AlexNet model.
 It can be observed that all three categories of hardware resources scale linearly as the number of computing units increases. The corresponding improvements on performance are also significant from using 2 CUs to 8 CUs. As system throughput continue to increase, memory bandwidth gradually reaches the onboard DRAM's limit (around 12.8GB/s), which introduces more frequent pipeline stalls resulting in a degraded performance gain. One can estimate from the reported data that the optimal parameters are \emph{VEC\_SIZE=8} and \emph{CU\_NUM=16} for the DE5-net board. Therefore, the shortest image classification time achieved are 43 ms for AlexNet and 718 ms for VGG-16 models, respectively. Fig.~\ref{fig:timeline} shows the profiled timeline for each kernels running these two CNNs models. Note that the final runtime without kernel profiling will be lower than that is shown in the figure.
To measure the power consumption, we blocked the power pins of the PCIe slot and powered the board through external port. The average power consumed by the board while running these two models are 27.3W and 29.8W, respectively.

We further compare the proposed design with other HSL-based designs in Table~\ref{table:result}. Since CNNs are multiplication-intensive, we adopt the number of DSPs consumed as the main factor to evaluate hardware resource utilizations. Our approach achieves a $34\%$ reduction on DSP resources while maintaining comparable performance with \cite{FPGA2016}. One could also estimate that the proposed architecture can obtain further improvements on performance over \cite{FPGA2016} when fixed-point data types were adopted. Moreover, the proposed design implements the full precision (32bit float format) CNN forward computation, which makes it also favorable to implement the backward propagation flow of model training.
 To make more straightforward comparison, we provide normalized performance as  "performance density" in the table. It is clear that our method outperforms previous works.

\begin{threeparttable}[t]
\caption{\normalsize Comparison with previous works.}
\label{table:result}
\begin{center}\small
\begin{tabular}{|c|c|c|c|}
\hline
& FPGA2016\cite{FPGA2016} & FPGA2015\cite{FPGA2015} & This work\\
\hline\hline
\multirow{2}{*}{Device} & Stratix-V & Virtex-7 & Stratix-V\\
 & GXA7 & VX485T & GXA7 \\
\hline
FPGA & 622K LUTs & 485K LUTs & 622K LUTs\\
Capacity & 256 DSP & 2800 DSP & 256 DSP\\
\hline
Design Scheme & OpenCL & Vivado HSL & OpenCL\\
\hline
Frequency & 120MHz & 100MHz & 181MHz\\
\hline
Precision & fixed(8-16b) & float & float\\
\hline
Classification & \multirow{2}{*}{45.7 ms\tnote{a}} & \multirow{2}{*}{21.6 ms\tnote{b}} & \multirow{2}{*}{43 ms\tnote{b}}\\
 Time & &  &  \\
\hline
Throughput & 31.8 GOPS\tnote{a} & 61.6 GOPS\tnote{b} & 33.9 GOPS\tnote{a} \\
\hline
DSP Consumed & 246 & 2240 & 162 \\
\hline
Performance & 0.13 & 0.027 & 0.21 \\
Density & GOPS/DSP & GOPS/DSP & GOPS/DSP \\
\hline
Power & 25.8W & 18.6W & 27.3W\\
\hline
\end{tabular}
\begin{tablenotes}
\footnotesize
\item[a] all operations for image classification.
\item[b] convolution operation only.
\end{tablenotes}
\end{center}
\end{threeparttable}

\section{Conclusion}
This work presents an open-source OpenCL-based FPGA accelerator for convolutional neural networks. A performance-cost scalable hardware architecture with efficiently pipelined kernels was proposed. Design spaces were explored by implementing two large-scale CNNs, AlexNet and VGG, on the DE5-net FPGA board. Results show that our scheme achieved significant improvements on performance density and resource utilizations compared to previous studies.





%

\end{document}